\documentclass{ws-procs9x6}

\begin{document}

\title{Electroweak and QCD Results from D\O}

\author{M.~EADS}

\address{Northern Illinois University,\\
E-mail: eads@fnal.gov}

\maketitle

\begin{center}
For the D\O~Collaboration
\end{center}

\abstracts{We present some of the results in the areas of QCD
and Electroweak physics for Run II of the D\O~experiment at
the Fermilab Tevatron. QCD results include dijet angular decorrelations
and inclusive jet and dijet cross sections. Electroweak results
include the decay of $Z$ bosons to tau pairs and several results
on gauge boson pairs. No deviations from the Standard Model have
been observed.}

\section{Introduction}

The D\O~experiment is a multi-purpose collider detector located at
the Fermilab Tevatron proton-antiproton collider. After undergoing
a substantial upgrade, the experiment is now taking data at 
$\sqrt{s} = 1.96~\mathrm{TeV}$. There are many new results 
based on this Run II data in the areas of Quantum Chromodynamics
(QCD) and Electroweak physics. 

\section{QCD Results}

%\subsection{Dijet Angular Decorrelations}

\begin{figure}[ht]
%\epsfxsize=10cm   %width of figure - will enlarge/reduce the figures
%\epsfbox{fig3.eps}
%\figurebox{2cm}{3cm}{} %to have a box alone 
\centerline{\epsfxsize=2.0in\epsfbox{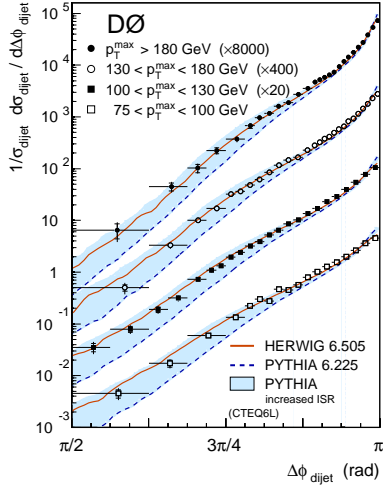}}   
\caption{The $\Delta \phi$ distributions in different
$p_T^{max}$ ranges. Results from HERWIG and PYTHIA are overlaid on
the data. HERWIG results agree with the data, but PYTHIA results
require modifications to the default parameters. \label{dijet_decorr_fig}}
\end{figure}

A study has been performed of the angular correlations in jets
produced in the D\O~detector \cite{dijet_decorr_ref}. At leading
order in QCD, jets are expected to be produced back-to-back 
in azimuth ($\phi$). One would then expect that the difference in
azimuthal angle between two jets is $\Delta \phi = \pi$. However,
higher order effects, such as additional soft radiation in the event,
will cause this angular difference to be less than $\pi$. The
distribution of $\Delta \phi$ is sensitive to higher order effects.
Additionally, this measurement does not rely on measuring the
energy of the jet and hence does not suffer from energy scale
systematic uncertainties. Figure \ref{dijet_decorr_fig} shows
the $\Delta \phi$ distribution for jets in four different 
transverse momentum bins compared to the predictions from Monte
Carlo generators.

%\subsection{Jet Cross Sections}

\begin{figure}[ht]
%\epsfxsize=10cm   %width of figure - will enlarge/reduce the figures
%\epsfbox{fig3.eps}
%\figurebox{2cm}{3cm}{} %to have a box alone 
\centerline{\epsfxsize=2.0in\epsfbox{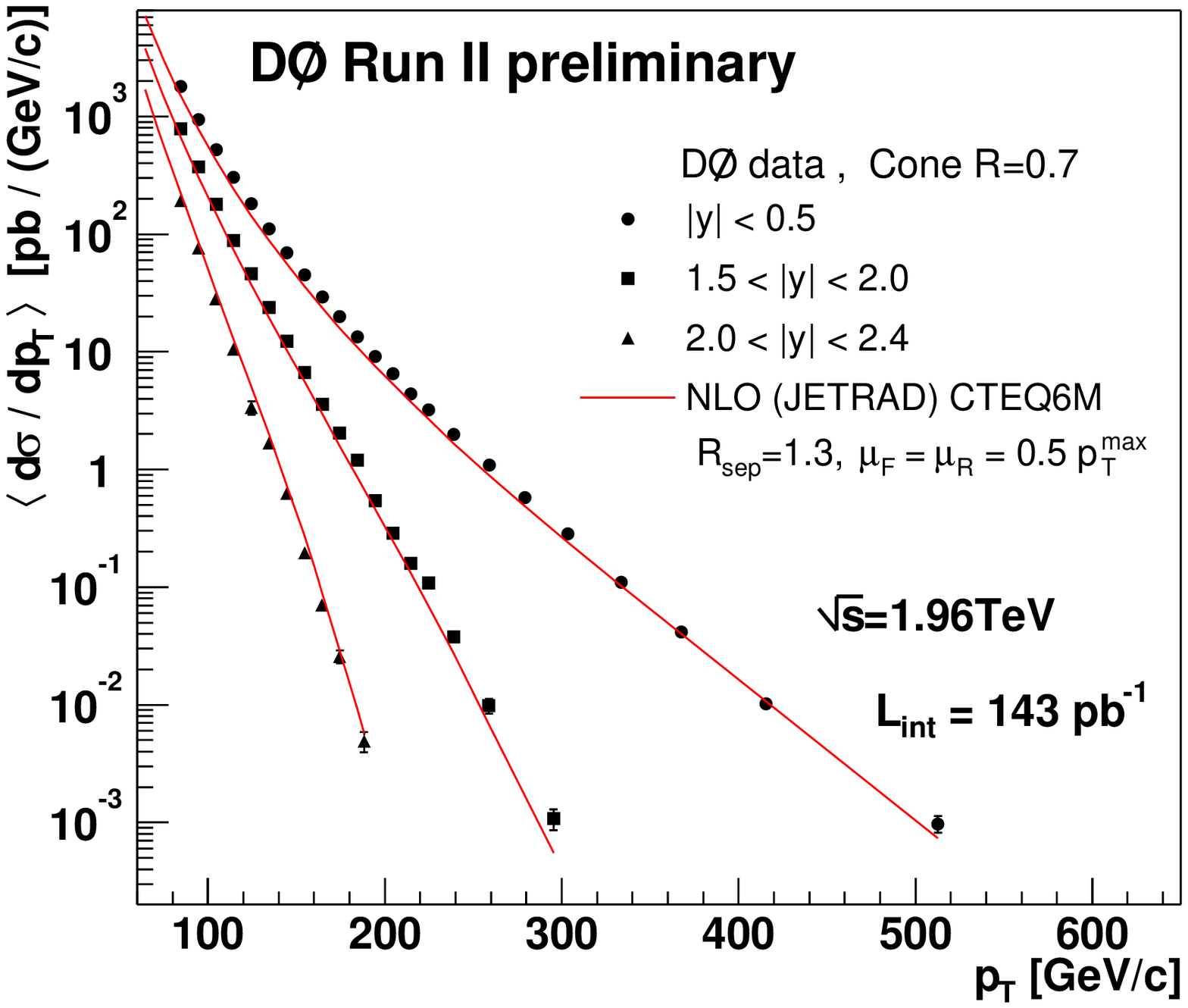}
	    \epsfxsize=2.0in\epsfbox{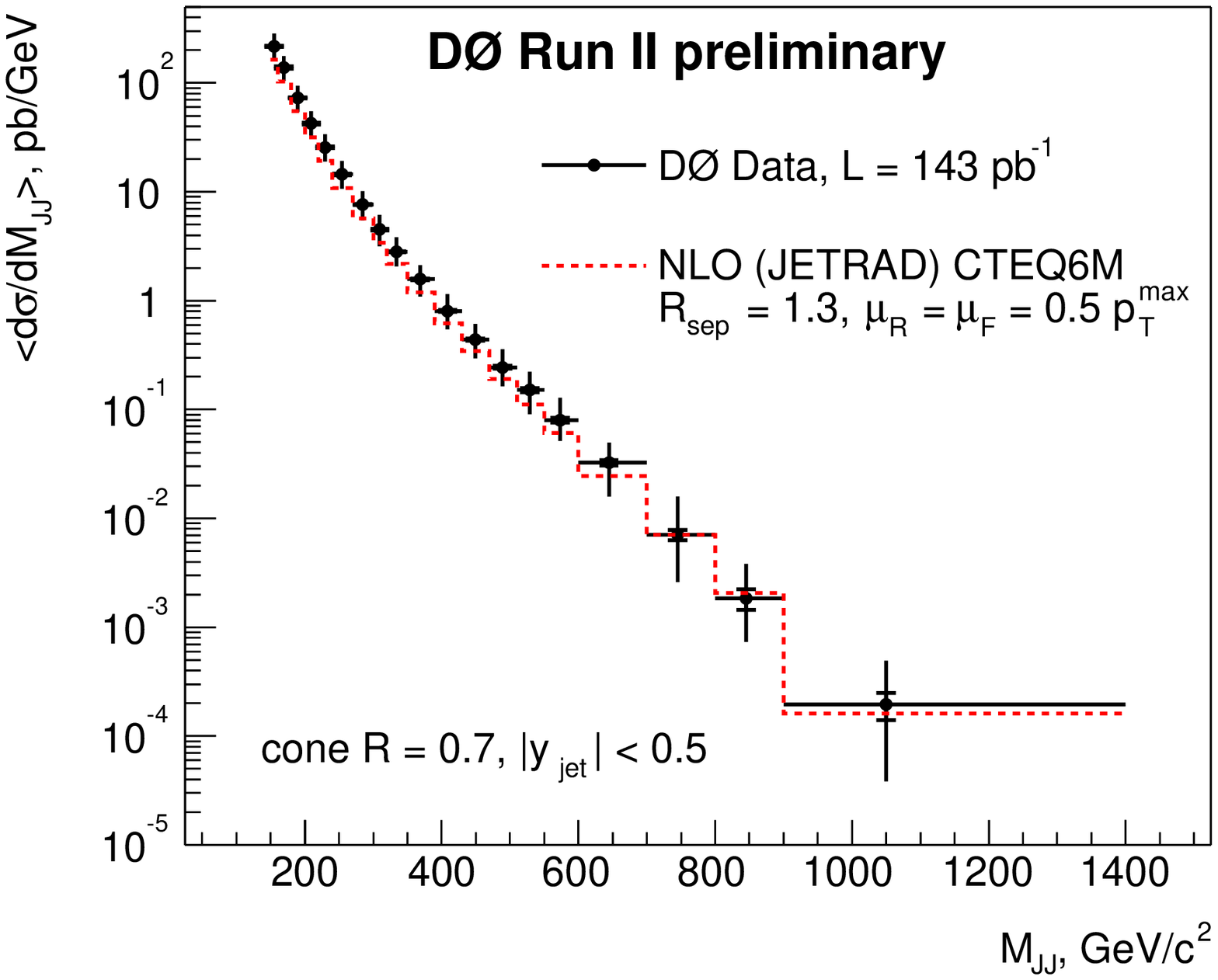}  } 
\caption{The left plot shows the inclusive jet cross section, 
measured in different
ranges of jet rapidity (with statistical errors only). NLO
pQCD calculations are overlaid on the data. 
The right plot shows the dijet cross section, 
measured at central rapidities. NLO pQCD calculations are
overlaid on the data. \label{jet_cross_fig}}
\end{figure}

The inclusive jet cross section and the dijet cross section have also been
measured. These distributions are sensitive to both the strong
coupling constant and to the parton density functions. Furthermore, 
many new physics models predict enhancements in the dijet mass cross
section at large values of invariant mass. Figure \ref{jet_cross_fig}
shows both the inclusive jet cross section and the dijet cross 
section distributions. These distributions are consistent with
the next-to-leading order perturbative QCD theoretical predictions.

\section{Electroweak Results}

D\O~has performed a measurement of the cross section times 
branching fraction for $Z$ bosons decaying to tau pairs
\cite{ztautau_ref}.
This measurement is a verification of the tau identification
abilities for the experiment as well as a test of lepton
universality. Furthermore, some physics models predict final
states that would result in an excess of tau pair events
over Standard Model predictions.

The analysis requires one tau to decay as a muon. The second tau
is identified using a neural network that has been trained using
the different tau decay topologies. 2008 candidate events are 
selected in 226 pb$^{-1}$ of data. Approximately  55\% of
these events are estimated to be background events. 
This results in a measurement
of $\sigma \cdot \mathrm{Br}(Z \to \tau \tau)
= 237 \pm 15_{stat} \pm 18_{sys} \pm 15_{lum}~\mathrm{pb}$. This
is consistent with the Standard Model prediction of $242 \pm 9 \mathrm{pb}$.

D\O~also has a variety of results on diboson production. 
At the Tevatron, pairs of gauge
bosons can be produced through $t$ or $u$ channel quark 
exchange, or can
be produced through an $s$-channel triple gauge boson vertex. The strength of
these triple gauge boson vertices is an important test of the
Standard Model. Diboson signatures are also important 
backgrounds for Higgs and new physics searches.

The cross section for the production of $p \overline{p} \to
W\gamma + X$ has been measured. This analysis requires the $W$ boson
to decay to either an electron or a muon and a neutrino. The
photon is identified by its signature in the calorimeter and the
absence of a matching track in the central tracking system. In 
the electron channel, 112 events are selected in 162 pb$^{-1}$ of data.
In the muon channel, 161 events are identified in 134 pb$^{-1}$ of data.
In both channels, the background is estimated to be approximately
half the number of events in data. The combination of both channels
results in a cross section measurement for $W\gamma X \to 
l\nu X$ of $14.8 \pm 1.6_{stat} \pm 1.0_{sys} \pm 1.0_{lum}~
\mathrm{pb}$. This is in agreement with the Standard Model prediction
of $16.0 \pm 0.4~\mathrm{pb}$.

\begin{figure}[ht]
%\epsfxsize=10cm   %width of figure - will enlarge/reduce the figures
%\epsfbox{fig3.eps}
%\figurebox{2cm}{3cm}{} %to have a box alone 
\centerline{\epsfxsize=2.0in\epsfbox{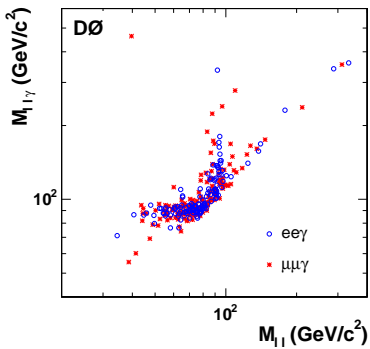}}   
\caption{Invariant mass of the dilepton system vs.
invariant mass of dilepton and a photon candidate. \label{zgamma_fig}}
\end{figure}

A measurement of the cross section times branching fraction 
has been performed for events
with a photon and a $Z$ boson, with the $Z$ boson decaying to 
electron or muon pairs \cite{zgamma_ref}. 
In the $ee\gamma$ channel, 33 data events are present
in 177 pb$^{-1}$ of data, with an estimated background of 
$4.7 \pm 0.7$ events. In the $\mu \mu \gamma$ channel, 68 data events
are present in 144 pb$^{-1}$ of data, with an estimated background of 
$10.1 \pm 1.3$ events. Figure \ref{zgamma_fig} shows the three body
invariant mass versus two body invariant mass for the candidate 
events. The combined cross section times branching fraction for both
channels is $3.90 \pm 0.51_{stat+sys} \pm 0.25_{lum}~\mathrm{pb}$,
which is in good agreement with the expected value of 4.3 pb.

A search for pairs of $W$ bosons has also been performed
\cite{ww_ref}. This 
analysis selects events with two opposite sign leptons
(electrons or muons) and missing energy. A total of 25 
events are selected in between 224 and 252 pb$^{-1}$ of data
(depending on the channel). The background has been estimated to be
$8.1 \pm 0.6_{stat} \pm 0.6_{sys} \pm 0.5_{lum}$. The data
represents a 5.2$\sigma$ excess over the background prediction.

\begin{figure}[ht]
%\epsfxsize=10cm   %width of figure - will enlarge/reduce the figures
%\epsfbox{fig3.eps}
%\figurebox{2cm}{3cm}{} %to have a box alone 
\centerline{\epsfxsize=2.0in\epsfbox{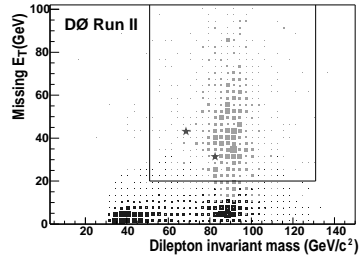}}   
\caption{Dilepton invariant mass vs. missing transverse energy
for expected $WZ \to \mu \mu \mu \nu$ events
(green or light grey) and for expected $Z$ + jet background
events (blue or dark grey). The central box shows the 
event selection criteria. \label{wz_fig}}
\end{figure}

A search for $WZ$ events has been performed by selecting 
events that have three leptons (electrons or muons) and 
missing energy \cite{wz_ref}. 
The missing transverse energy versus
dilepton invariant mass distribution is shown in Figure
\ref{wz_fig}. A total of three candidate events are observed
in data in approximately 300 pb$^{-1}$ of data. There are
$2.04 \pm 0.13$ expected signal events and $0.71 \pm 0.08$ expected
background events. A 95\% confidence level limit on the cross 
section has been set at 13.3 pb.

\begin{figure}[ht]
%\epsfxsize=10cm   %width of figure - will enlarge/reduce the figures
%\epsfbox{fig3.eps}
%\figurebox{2cm}{3cm}{} %to have a box alone 
\centerline{\epsfxsize=2.0in\epsfbox{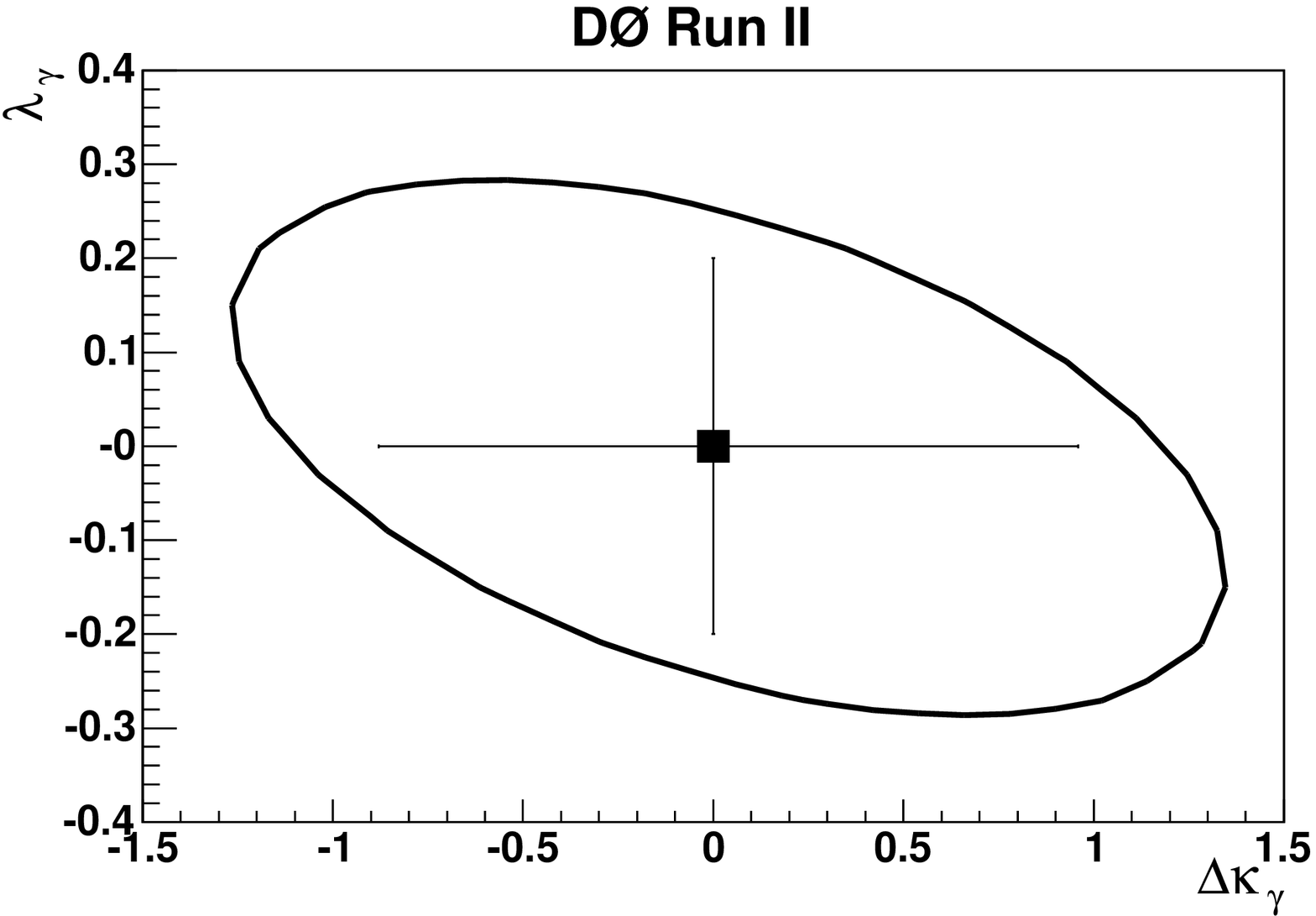}   
            \epsfxsize=2.0in\epsfbox{dzero_ew_qcd_fig5a.eps}}   
\caption{Left plot shows the 
limits on the $WW\gamma$ coupling parameters
$\Delta \kappa_\gamma$ and $\lambda_\gamma$. The
point indicates the SM value with the error bars showing the
95\% CL intervals in one dimension. the ellipse represents
the two-dimensional 95\% CL exclusion contour.
The right plot shows
two-dimensional coupling limits (inner contour) on
$\lambda_z$ vs. $\Delta g_1^z$ at 95\% C.L. for $\Lambda =
1.5~\mathrm{TeV}$. The outer contour is the limit from
$S$-matrix unitarity.
 \label{anom_fig}}
\end{figure}

Limits have also been set on anomalous $WW\gamma$ and $WWZ$
couplings. These couplings can be parameterized in an
effective Lagrangian, with a scale factor $\Lambda$.
Figure \ref{anom_fig} shows the limits set on the
anomalous coupling parameters. All of these anomalous coupling
limit parameters are predicted to be zero in the Standard
Model.

\end{document}